\documentstyle[aps,prb,twocolumn,epsf,floats]{revtex}
\begin{document}

\ifx\hyperlink\hyperudefined
\else
\errmessage{Looks ugly with hyperlinks.}
\stop
\fi

\draft

\title{Effects of anharmonic strain on phase stability of epitaxial
films and superlattices: applications to noble metals}
\author{V. Ozoli\c{n}\v{s}, C. Wolverton, and Alex Zunger}
\address{National Renewable Energy Laboratory, Golden, CO 80401}
\date{September 12, 1997}
\maketitle

{\let\clearpage\relax
\twocolumn[
\widetext\leftskip=0.1075\textwidth \rightskip\leftskip
\begin{abstract}
Epitaxial strain energies of epitaxial films and bulk superlattices
are studied via first-principles total energy calculations using the
local-density approximation. Anharmonic effects due to large lattice
mismatch, beyond the reach of the harmonic elasticity theory, are
found to be very important in Cu/Au (lattice mismatch 12\%), Cu/Ag
(12\%) and Ni/Au (15\%). We find that $\langle 001 \rangle$ 
is the elastically soft direction for biaxial expansion of Cu and Ni, but 
it is $\langle 201 \rangle$ for large biaxial compression
of Cu, Ag, and Au. The stability of superlattices is discussed in
terms of the coherency strain and interfacial energies. 
We find that in phase-separating systems such as Cu-Ag the
superlattice formation energies {\it decrease\/} with superlattice
period, and the interfacial energy is positive. Superlattices are
formed easiest on (001) and hardest on (111) substrates. For ordering
systems, such as Cu-Au and Ag-Au, the formation energy of
superlattices {\it increases\/} with period, and interfacial energies
are {\it negative.\/} These superlattices are formed easiest on (001)
or (110) and hardest on (111) substrates. For Ni-Au we find a hybrid
behavior: superlattices along $\langle 111 \rangle$ and 
$\langle 001 \rangle$ behave like in phase-separating systems, while
for $\langle 110 \rangle$ they behave like in ordering systems.
Finally, recent experimental results on epitaxial
stabilization of disordered Ni-Au and Cu-Ag alloys, immiscible in the
bulk form, are explained in terms of destabilization of the phase
separated state due to lattice mismatch between the substrate and
constituents.
\end{abstract}
]}

\narrowtext

\section{Introduction}
\label{sec:intro}

Recently, there has been much
interest\cite{oneill90,mueller91,brune94,gunther95,hwang95,potschke91,vidali96,ramirez,bauer86,merwe91,gautier91,mottet92,hamilton95,zwlu-SLs}
in growth of epitaxial metal films and superlattices due to their
unusual physical properties. The quality and structure of these
systems is of paramount importance for applications. Epitaxial
monolayer and multilayer (up to 10 layers) formation has been observed
for many metal/semiconductor and metal/metal combinations. Most
metal/metal superlattices have been grown for elements in different
crystal structures (e.g., fcc/bcc) and with considerable size mismatch
(e.g., 10\% for Cu/Nb\cite{lowe81,banerjee,chun84}). Furthermore,
elemental metals and alloys have been found to form epitaxially in
structures which are unstable in bulk
form.\cite{epi_review,wang94,saleh,kim96,worm}
Recently, the topic of surface alloy formation in bulk immiscible
systems has attracted considerable
attention.\cite{mcrae86,liu91,chan92,rousset92,chamb92,chamb93,roder93,tzeng93,danes,danes2,boerma94,altman94,nagl95,stevens95,schmid96,adams96,johnson88,bozzolo95,tersoff95,danes97}
These systems are usually strained due to film/substrate lattice
mismatch. One would like to understand and predict the
stability of these types of strained materials.
In order to do so, one requires knowledge of two types of energies.
The stability of epitaxial $A_{1-x}B_x$ alloy films and strained 
$A_pB_q$ superlattices depends on (i) the energies of coherently strained
{\it constituents\/} $A$ and $B$, and (ii) the formation energy of
$A_{1-x}B_x$ or $A_pB_q$ itself. Regarding (i), previous theoretical
studies\cite{epi_review,HB78,yang94,marcus95,japanese,cahn,larche85,cremoux82,flynn86,stringfellow,chiang89,dmwood88,dmwood92,russians,dblaks92}
have described these energies using
harmonic models, but we are interested here in large
strains for which the harmonic theory could break down.  Thus,
we develop a generalization of previous methods to treat the
anharmonic epitaxial strain energies of the constituents.
Regarding (ii), these energies depend on the configuration
degrees of freedom of the epitaxial film, so their calculation
requires statistical methods.\cite{dblaks92,NATO} 
In the present paper we investigate items (i) and (ii) above
using accurate first-principles LDA calculations. 

As for (i), the constituent strain energy, we
find that the harmonic strain theory,\cite{epi_review,japanese} 
predicting a single, universal relation for elastically soft
directions, breaks down for sufficiently large substrate/film
lattice mismatch.
We find that under biaxial {\it expansion,\/} noble metals are soft
along $\langle 001 \rangle$, but that under {\it compression\/} the
soft direction changes to $\langle 201 \rangle$.
It is shown
that the softness of $\langle 001 \rangle$ is a consequence
of low bcc/fcc energy differences in noble metals, while
the softness of $\langle 201 \rangle$ under compressive strain 
can be explained by loose packing of atoms in the \{201\} planes.
Furthermore, the elastic strain energy 
as a function of direction exhibits qualitative 
shifts in the hard and soft strain directions, which cannot be
guessed from the harmonic elasticity theory. For instance,
we find that $\langle 110 \rangle$ becomes the hardest direction 
under biaxial expansion, and $\langle 201 \rangle$ becomes
the softest direction under biaxial compression, while
the harmonic theory always predicts either $\langle 111 \rangle$
as the hardest and $\langle 001 \rangle$ as the softest direction, or
vice versa. 

Regarding (ii), the formation energy, we find that
the anomalous elastic softness of the constituents along 
$\langle 001 \rangle$ and $\langle 201 \rangle$ leads to low 
constituent strain energy in superlattices along these directions,
which makes them more stable than superlattices along other
$\widehat{G}$. For instance, in the size-mismatched systems
Cu-Au, Cu-Ag, and Ni-Au, $A_n B_n$ superlattices along
$\langle 001 \rangle$  are the most stable for all periods $n$.
Interfacial energies are found to be negative in Ag-Au and Cu-Au
(reflecting their bulk miscibility), and positive in the phase
separating systems Cu-Ag and Ni-Au. However, attraction between
(110) interfaces in Ni-Au is very strong and favors short-period
($n \propto 2$) superlattices over long-period superlattices with
few interfaces. 

In the case
of epitaxially grown disordered alloys, we find that the biaxial
constraint on the phase separated constituents may stabilize the alloy
with respect to phase separation. The stabilization effect is always 
greater on substrates oriented along elastically hard directions
(i.e., with high constituent strain energy) like $\langle 111 \rangle$
than along soft directions like $\langle 001 \rangle$. For instance,
on lattice-matched substrates, epitaxial Ni$_{0.5}$Au$_{0.5}$ alloys
are stable at all temperatures, and Cu$_{0.5}$Ag$_{0.5}$ alloys are
stable for $T>150$~K if grown on a (111) substrate,
although both these systems phase separate in bulk form or
if grown on a (001) substrate.
These predictions agree very well with recent experimental
observations.\cite{danes,stevens95}

\section{Bulk and epitaxial stability criteria}
\label{sec:energetics}

The stability of either free-standing or coherently strained alloys
and superlattices requires specification of
(i) epitaxial strain energies of pure constituents due to
film/substrate lattice mismatch, (ii) formation enthalpies of
disordered alloys (with respect to either strained or unstrained bulk
constituents) and superlattices.
In this section, we define these quantities and discuss the physical
situations where they should be used.

\subsection{Epitaxial strain energies of elemental constituents}

We start by considering (i) above, which is a common element to
alloys and superlattices.
Consider a film of pure element $A$ coherently strained on a substrate
oriented along direction $\widehat{G}$ with surface unit cell 
vectors ${\bf a}_1$ and ${\bf a}_2$, orthogonal to $\widehat{G}$.
We assume that the film, being much thinner than 
the substrate, maintains coherency with the substrate and plastically
deforms to accomodate the lattice mismatch at the interface.
This assumption is valid for films thinner than the critical
thickness for the nucleation of misfit dislocations.
Furthermore, we consider
films which are thick enough so that the chemical interaction energy
at the film/substrate interface and film/vacuum surface is negligibly
small in comparison with the elastic 
deformation energy of the film. Under these assumptions, the {\it 
epitaxial strain energy\/} 
$\Delta E_A^{\rm epi} ({\bf a}_1,{\bf a}_2,\widehat{G})$ of film $A$
is the strain energy of element $A$ deformed in the growth plane
to the unit cell vectors $\{ {\bf a}_1, {\bf a}_2 \}$ of the
substrate, and relaxed with respect to the out-of-plane 
vector ${\bf c}$:
\begin{equation}
\label{eq:elementepi}
\Delta E_A^{\rm epi} ({\bf a}_1,{\bf a}_2,\widehat{G}) =
\min_{\bf c} \left[ E^{\rm tot}_A ({\bf a}_1,{\bf a}_2,{\bf c})
\right] - E^{\rm tot}_A (a_A).
\end{equation}
In what follows, we are interested in the case where both the 
substrate and the unstrained bulk element $A$ have the fcc
crystal lattice. Then ${\bf a}_1$ and ${\bf a}_2$ are 
proportional to the equilibrium unstrained lattice vectors
of fcc $A$, ${\bf a}^)0_i (A)$:
\begin{equation}
\label{eq:scaling}
{\bf a}_i = \left(\frac{a_s}{a_A}\right) {\bf a}_i^0(A),\;\;i=1,2,
\end{equation}
where $a_s$ and $a_A$ are fcc lattice parameters of the substrate and
$A$, correspondingly.
The epitaxial strain energy becomes a function of the substrate
lattice constant and direction $\widehat{G}$ only:
\begin{equation}
\label{eq:simpleepi}
\Delta E_A^{\rm epi} [(a_s/a_A) {\bf a}_1, (a_s/a_A) {\bf
a}_2,\widehat{G}] \equiv \Delta E^{\rm epi}_A (a_s,\widehat{G}).
\end{equation}
LDA calculations of $\Delta E^{\rm epi}_A (a_s,\widehat{G})$ are
described in Sec.~\ref{sec:epitaxy}.

\subsection{Formation enthalpies of alloys and superlattices}

Like the formation enthalpy of any ordered bulk compound, the 
{\it formation enthalpy\/} 
$\Delta H_{\rm SL}^{\rm bulk} (pq, \widehat{G})$ of an
$A_pB_q$ unstrained (bulk) superlattice is defined as the energy gain
or loss with respect to {\it unstrained\/} bulk constituents:
\begin{eqnarray}
\nonumber
\lefteqn{\Delta H_{\rm SL}^{\rm bulk} (pq, \widehat{G}) = 
E^{\rm tot} (A_pB_q, \widehat{G})} \\
\label{eq:bulkH}
	& - & \left[ \frac{p}{p+q} E^{\rm tot}_A(a_A) + 
		\frac{q}{p+q} E^{\rm tot}_B (a_B) \right],
\end{eqnarray}
where $a_A$ is the equilibrium lattice constant of the unstrained bulk
element $A$ and $E_A^{\rm tot}(a_A)$ is the total energy of $A$. 
This enthalpy characterizes the propensity to form superlattices with
respect to the phase separated bulk constituents.
If $\Delta H_{\rm SL}^{\rm bulk} (pq,\widehat{G})<0$, the unstrained
superlattices are energetically favored over the phase separation,
while the phase separated state is favored if 
$\Delta H_{\rm SL}^{\rm bulk} (pq,\widehat{G})>0$.
To be stable, free-standing bulk superlattices must satisfy
stability criteria with respect to at least: 
(i) phase separation into unstrained bulk
constituents and (ii) formation of a configurationally disordered
bulk alloy. The {\it bulk mixing enthalpy\/},
$\Delta H_{\rm mix}^{\rm bulk} (A_{1-x}B_x)$, 
of the alloy is given by: 
\begin{eqnarray}
\nonumber
\lefteqn{\Delta H_{\rm mix}^{\rm bulk} (A_{1-x}B_x) = 
E^{\rm tot} (A_{1-x}B_x)} \\
\label{eq:Hmix}
	&& -\left[ (1-x) E^{\rm tot}_A(a_A) + 
		x E^{\rm tot}_B (a_B) \right],
\end{eqnarray}
where $x=q/(p+q)$ is the composition and 
$E^{\rm tot} (A_{1-x}B_x)$ is the total energy per atom of the
configurationally random alloy.

If $\Delta H_{\rm mix}^{\rm bulk} (A_{1-x}B_x) < 
\Delta H^{\rm bulk}_{\rm SL} (A_pB_q) < 0$, then 
both the superlattice and disordered alloy are stable with respect to
phase separation, but the superlattice is unstable with
respect to disordering. However, if
$\Delta H^{\rm bulk}_{\rm SL} (A_pB_q) < 
\Delta H_{\rm mix}^{\rm bulk} (A_{1-x}B_x) < 0$, then
superlattices are stable with respect to both phase separation and
disordering, and it may be possible to grow them. 

The bulk formation enthalpy of a superlattice,
$\Delta H^{\rm bulk}_{\rm SL} (pq,\widehat{G})$, can be
separated into two components. To identify them, it is useful to first
consider the infinite period superlattice limit
$p,q \rightarrow \infty$, where $A/B$ interfacial
interactions contribute a negligible amount of order ${\cal
O}(1/p)$. In this case, the bulk formation enthalpy of 
$A_\infty B_\infty$ superlattice is given by
\begin{eqnarray}
\label{eq:ECSdef}
\lefteqn{\Delta H_{\rm SL}^{\rm bulk} (pq \rightarrow \infty,
\widehat{G}) \equiv \Delta E_{\rm CS}^{\rm eq} (x,\widehat{G})} \\
\nonumber
 &=& \min_{{\bf a}_1,{\bf a}_2} \left[ 
(1-x) \Delta E_A^{\rm epi} ({\bf a}_1,{\bf a}_2,\widehat{G}) +
x \Delta E_B^{\rm epi} ({\bf a}_1,{\bf a}_2,\widehat{G}) \right],
\end{eqnarray}
where $\Delta E_A^{\rm epi}$ is
the epitaxial deformation energy of $A$, given by
Eq.~(\ref{eq:elementepi}). We define this energy as the ``constituent
strain'' (CS) to emphasize that in this limit the superlattice
formation enthalpy depends only on its strained constituents.
This is also the energy required to keep $A$ and $B$ coherent.

For finite-period superlattices, the formation energy is determined not
only by the elastic strain energy, but also by
interactions between unlike atoms at $A/B$ interfaces.
We define this {\it interfacial energy\/} $I(pq,\widehat{G})$ as: 
\begin{equation}
\label{eq:Eint_general}
\Delta H_{\rm SL}^{\rm bulk}(pq, \widehat{G}) - 
\Delta H_{\rm SL}^{\rm bulk} (pq \rightarrow \infty, \widehat{G})
\equiv \frac{4}{p+q} I(pq, \widehat{G}).
\end{equation}
It is the total energy per layer
of a single interface between infinite slabs of $A$ and $B$ oriented
along $\widehat{G}$.
$I(\infty)<0$ signals that the interface is energetically favored, while
$I(\infty)>0$ indicates that an isolated interface is not preferred,
and long-period superlattices with fewer interfaces are usually more
stable than the short-period ones (however, this simple argument is
not always true, see the following discussion).

For equiatomic $(A)_n/(B)_n$ superlattices Eq.~(\ref{eq:Eint_general})
becomes:
\begin{equation}
\label{eq:Einterface}
\Delta H_{\rm SL}^{\rm bulk}(n, \widehat{G}) = 
\frac{2I(n,\widehat{G})}{n} + 
\Delta E_{\rm CS}^{\rm eq} (x=0.5,\widehat{G}).
\end{equation}
For small $n$ interfaces will interact with each other. We describe
this process by the interface interaction energy 
$\delta I(n,\widehat{G})$:
\begin{equation}
\label{eq:Jinterf(n)}
\delta I(n,\widehat{G}) = I(n, \widehat{G}) -
	I(n \rightarrow \infty, \widehat{G}).
\end{equation}
Negative $\delta I(n,\widehat{G})$ may favor short-period superlattices
over long-period superlattices even if the interfacial
energy $I(n \rightarrow \infty, \widehat{G})$ is positive. For this to
happen it is necessary that
\begin{equation}
\label{eq:J1}
\delta I(n,\widehat{G}) < - | I(n \rightarrow \infty, \widehat{G}) |
\end{equation}
In Sec.~\ref{sec:finiteSL} we show that this unusual phenomenon 
occurs in Ni-Au.

If a disordered alloy is grown epitaxially on a 
{\it lattice-matched\/} fcc substrate, its stability with respect to
phase separation is given by the {\it epitaxial mixing enthalpy\/}:
\begin{eqnarray}
\nonumber 
\lefteqn{\delta H_{\rm mix}^{\rm epi} (A_{1-x}B_x) = 
\Delta H_{\rm mix}^{\rm bulk} (A_{1-x}B_x)} \\
\label{eq:epitaxialHmix}
	&&-(1-x) \Delta E^{\rm epi}_A (a_s,\widehat{G}) -
	x \Delta E^{\rm epi}_B (a_s,\widehat{G}),
\end{eqnarray}
where $\Delta E^{\rm epi}_A(a_s,\widehat{G})$ is the epitaxial
strain energy of Eq.~(\ref{eq:simpleepi}), accounting for the fact
that the phase-separated consituents must also 
be lattice-matched with the substrate.
Due to the presence of these terms, disordered alloys
may form epitaxially 
[$\delta H_{\rm mix}^{\rm epi} (A_{1-x}B_x) < 0$]
even if the corresponding bulk alloys phase separate
[$\Delta H_{\rm mix}^{\rm bulk} (A_{1-x}B_x) > 0$].
This situation is especially likely to occur for elastically hard
directions $\widehat{G}$ with large values of
$\Delta E^{\rm epi}_{A,B} (a_s,\widehat{G})$, for instance
$\langle 111 \rangle$ and $\langle 110 \rangle$ 
(see Sec~\ref{sec:qs}).

The objective of this work is to calculate
$\Delta H^{\rm bulk}_{\rm SL} (A_pB_q)$ [Eq.~(\ref{eq:bulkH})],
$\Delta H_{\rm mix}^{\rm bulk} (A_{1-x}B_x)$ [Eq.~(\ref{eq:Hmix})] and
$\delta H_{\rm mix}^{\rm epi} (A_{1-x}B_x)$
[Eq.~(\ref{eq:epitaxialHmix})] from first principles for
Ag-Au, Cu-Ag, Cu-Au, and Ni-Au.
This requires:

(a) Epitaxial strain energies of pure constituents,
$\Delta E^{\rm epi}_A (a_s,\widehat{G})$  [Eq.~(\ref{eq:simpleepi})],
for Ag, Au, Cu and Ni. This is described in Sec.~\ref{sec:epitaxy}.

(b) Equilibrium constituent strain energy
$\Delta E_{\rm CS}^{\rm eq}$ [Eq.~(\ref{eq:ECSdef})] for
Ag-Au, Cu-Ag, Cu-Au, and Ni-Au. This is
described in Sec.~\ref{sec:Hinfinite}.

(c) The interfacial energy 
$I(pq,\widehat{G})$ of Eq.~(\ref{eq:Einterface}) requires
$\Delta H^{\rm bulk}_{\rm form} (A_p B_q, \widehat{G})$
for arbitrary $pq$ and $\widehat{G}$.
$\Delta H^{\rm bulk}_{\rm mix} (A_{1-x} B_x)$
and $\delta H^{\rm epi}_{\rm mix} (A_{1-x} B_x)$ require the
total energy of a configurationally disordered solid solution.
All these quantities are obtained from the mixed-space cluster
expansion as described in Sec.~\ref{sec:CE}.

\section{Elemental epitaxial films}
\label{sec:epitaxy}

\subsection{Anharmonic epitaxial strain in thin films of pure
elements: Analytic forms}
\label{sec:epi_theory}

The epitaxial strain energy [Eq.~(\ref{eq:simpleepi})] of a film of
element $A$ (with an equilibrium fcc lattice constant $a_A$) on a fcc
substrate with lattice constant $a_s$, oriented along direction
$\widehat{G}$, is conveniently obtained in a two-step process
considered by Hornstra and Bartels.\cite{HB78} 
First, the fcc crystal of bulk $A$ is uniformly
stretched (or compressed) to the lattice constant of the substrate
$a_s$. The energy change relative to free $A$ is given by the
hydrostatic bulk deformation energy $\Delta E^{\rm bulk}_A (a_s)$. In
the second step, out-of-plane unit cell vector ${\bf c}$ of the
film relaxes to satisfy Eq.~(\ref{eq:elementepi}).
The change $\Delta {\bf c} = {\bf c} - (a_s/a_A) {\bf c}^0$
(where ${\bf c}^0$ is the fcc lattice vector of unstrained $A$),
has components parallel [$\Delta {\bf c}_\parallel$] and
perpendicular [${\Delta \bf c}_\perp$] to the growth direction
$\widehat{G}$. The parallel component $\Delta {\bf c}_\parallel$
changes the volume of the unit cell and thus has a large effect on the
total energy. In contrast, the so-called shear strain 
$\Delta {\rm c}_\perp$ shifts planes orthogonal to $\widehat{G}$
and does not change the volume of the unit cell. Consequently, it has
a much smaller effect on the total energy. Furthermore, this strain
vanishes by symmetry for directions $\langle 001 \rangle$, 
$\langle 111 \rangle$ and $\langle 110 \rangle$,
and the shear strain energy must have zero angular derivatives 
at these points. Therefore, we neglect the shear strain $\Delta
c_\perp$ also for low-symmetry directions. 
Bottomley and Fons\cite{japanese} have shown that this approximation
introduces rather small errors in the {\it harmonic\/} epitaxial strain
energies.

Neglecting the shear strain $\Delta {\bf c}_\perp$, the strain
energy of element $A$ is then a function of the direction
$\widehat{G}$ and two scalar variables, $a_s$ and 
$\epsilon_\parallel = |\Delta {\bf c}_\parallel | / a_s - 1$.
The epitaxial strain energy
$\Delta E^{\rm epi}_A (a_s,\widehat{G})$ of Eq.~(\ref{eq:simpleepi}) 
is the minimum of the strain energy with respect to
$\epsilon_\parallel$ at a fixed substrate lattice constant $a_s$:
\begin{equation}
\label{eq:simple_strainE}
\Delta E_A^{\rm epi} (a_s,\widehat{G}) = \min_{\epsilon_\parallel}
\left[ E^{\rm tot}_A (a_s, \epsilon_\parallel,\widehat{G}) \right]
- E_A^{\rm tot} (a_A).
\end{equation}
The epitaxial strain energy $\Delta E^{\rm epi}_A (a_s,\widehat{G})$
is related to the epitaxial softening function\cite{dmwood88,epi_review}
$q(a_s,\widehat{G})$ by the relation:
\begin{equation}
\label{eq:qdef}
q(a_s, \widehat{G}) = \frac{\Delta E_A^{\rm epi} (a_s,\widehat{G})}
{\Delta E_A^{\rm bulk} (a_s)},
\end{equation}
where $\Delta E_A^{\rm bulk} (a_s)$ is the hydrostatic deformation
energy of fcc $A$ to the substrate lattice constant $a_s$.
The function Eq.~(\ref{eq:qdef}) quantifies energy lowering due 
to the relaxation of ${\bf c}(A)$ in the second step of the
deformation process considered above.

\begin{figure}
\epsfxsize=3in
\centerline{\epsffile{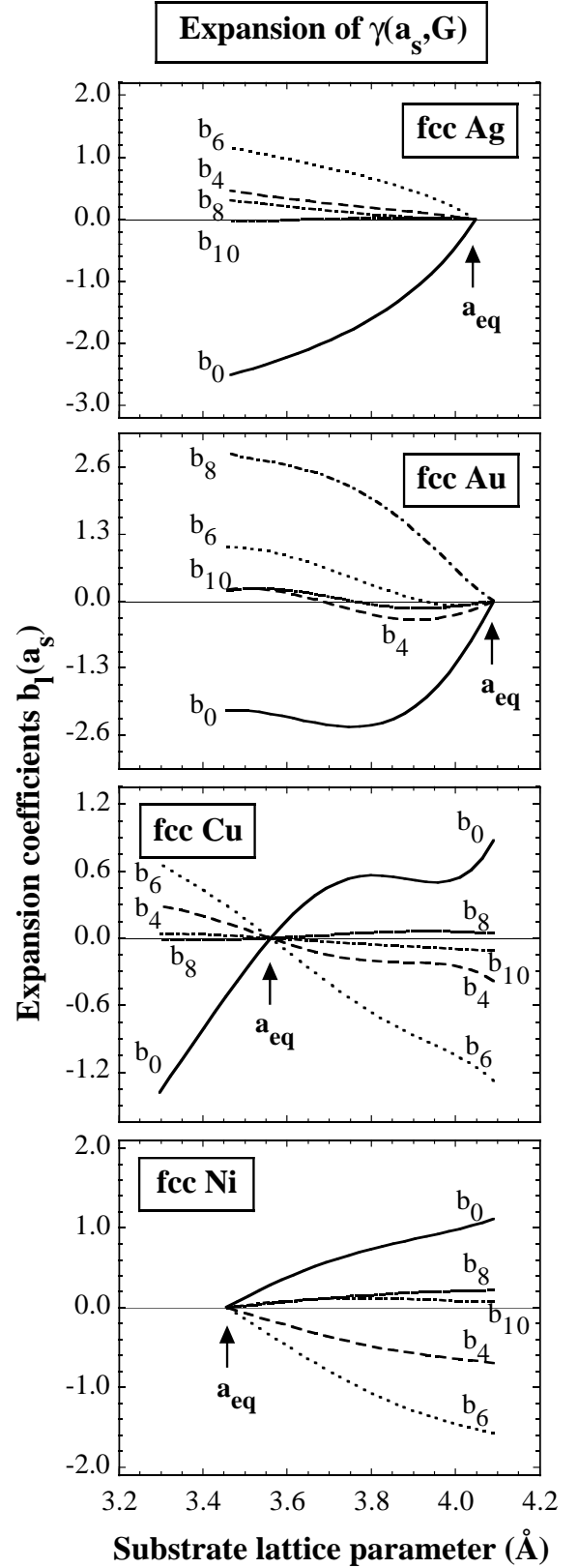}}
\vskip 5mm
\caption{Expansion coefficients $b_l(a_s)$ of
Eq.~(\protect\ref{eq:gamma_general}) for Ag, Au, Cu, and Ni.}
\label{fig:bl}
\end{figure}

The harmonic elasticity theory without the shear strain
gives\cite{epi_review,dblaks92,japanese}
$q_{\rm harm} (\widehat{G})$ which depends on the growth direction
$\widehat{G}$ but not on the substrate lattice constant $a_s$:
\begin{equation}
\label{eq:qharm}
q_{\rm harm} (\widehat{G}) = 1 - \frac{B}{C_{11} +
\Delta \; \gamma_{\rm harm} (\widehat{G})},
\end{equation}
where $B = \frac{1}{3} (C_{11} + 2 C_{12})$ is the bulk modulus, 
$\Delta = C_{44} - \frac{1}{2} (C_{11}-C_{12})$ is the elastic
anisotropy parameter, and
$\gamma_{\rm harm} (\widehat{G})$ is a geometric function
of the spherical angles formed by $\widehat{G}$:
\begin{eqnarray}
\nonumber
\lefteqn{\gamma_{\rm harm} (\phi, \theta) = 
\sin^2 (2\theta) + \sin^4 (\theta) \sin^2 (2\phi)} \\
\label{eq:gamma_harm}
&=& \frac{4}{5} \sqrt{4\pi} [K_0(\phi,\theta) - \frac{2}{\sqrt{21}}
K_4(\phi,\theta) ].
\end{eqnarray}
$K_l$ is the Kubic harmonic of angular momentum $l$.
The equilibrium value of the $c_\parallel/a$ ratio of the film is
given by 
\begin{equation}
\label{eq:c/a}
c^{\rm eq}_\parallel (a_s,\widehat{G}) =  a_s (1+\epsilon_\parallel) = 
a_A - [2-3q_{\rm harm} (\widehat{G})] (a_s - a_A).
\end{equation}
For the principle high-symmetry directions we have
\begin{equation}
\gamma_{\rm harm}([001]) =0, \;\; 
\gamma_{\rm harm}([110]) = 1, \;\; 
\gamma_{\rm harm}([111]) = \frac{4}{3}.
\end{equation}
A parametric plot of $\gamma$ is presented in
Ref.~\onlinecite{dblaks92}, which shows that the minimum of
$\gamma(\widehat{G})$ is along $\langle 001 \rangle$ and the 
maximum -- along $\langle 111 \rangle$.
Therefore, depending on the sign of the elastic anisotropy $\Delta$,
$q_{\rm harm} (\widehat{G})$ is either lowest for the
$\langle 001 \rangle$ direction, and then $q_{\rm harm}([111])$ 
is the highest, or vice versa.
Other directions always have intermediate values of 
$q_{\rm harm}(\widehat{G})$. 

\begin{figure}
\epsfxsize=2.6in
\centerline{\epsffile{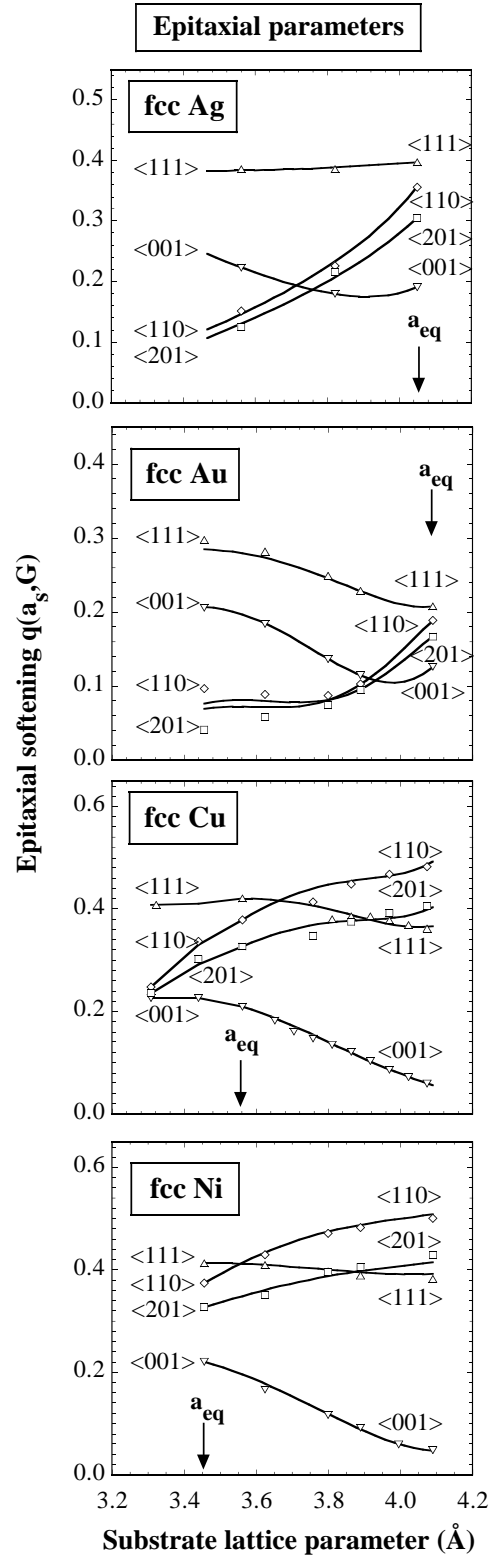}}
\vskip 5mm
\caption{The calculated epitaxial softening functions
$q(a_s,\widehat{G})$ for Cu, Ni, Ag and Au. Points represent the
directly calculated LDA values and lines show the fit
using
Eqs.~(\protect\ref{eq:q_general})--(\protect\ref{eq:gamma_general}).}
\label{fig:qs}
\end{figure}

If anharmonic effects are important, $q$ becomes a function
of the substrate lattice parameter $a_s$.
As we will show in Sec.~\ref{sec:qs}, for deformations
$2(a_s-a_A)/(a_s+a_A)$ of approximately  $4\%$,
the ``exact'' LDA $q(a_s, \widehat{G})$ exhibits appreciable
dependence on the substrate lattice parameter $a_s$ and certain
qualitative features cannot be reproduced by the harmonic functional
form Eqs.~(\ref{eq:qharm})--(\ref{eq:gamma_harm}).
Furthermore, sufficiently large epitaxial strains may
take the lattice from the face-centered cubic (fcc) structure 
into other low-energy structures [e.g., 
body-centered cubic (bcc) and body-centered tetragonal (bct)],
causing anomalous softening of $q(a_s,\widehat{G})$ for these
directions. Section~\ref{sec:qs} shows that this indeed happens for
$\langle 001 \rangle$ epitaxial strain when $a_s > a_A$.
Therefore, Eqs.~(\ref{eq:qharm})--(\ref{eq:gamma_harm})
must be generalized to account for nonlinear effects beyond the reach
of the harmonic theory. This is achieved by replacing in
Eq.~(\ref{eq:qharm}) $\gamma_{\rm harm} (\widehat{G})$ by 
$\gamma (a_s, \widehat{G})$, where
\begin{equation}
\label{eq:gamma_general}
\gamma(a_s,\widehat{G}) = \gamma_{\rm harm} (\widehat{G}) + 
\sum_{l=0}^{l_{\rm max}} b_l(a_s) \, K_l (\widehat{G})
\end{equation}
includes higher Kubic harmonics. For cubic systems $l=0,4,6,8, \ldots$
The general expression for $q$ is
\begin{equation}
\label{eq:q_general}
q (a_s, \widehat{G}) = 1 - \frac{B}{C_{11} +
\Delta \; \gamma (a_s, \widehat{G})}.
\end{equation}
We have chosen this particular form for $\gamma$ since it guarantees 
that all expansion coefficients tend to zero in the harmonic limit:
\begin{equation}
\lim_{a_s \rightarrow a_A} b_l (a_s) = 0.
\end{equation}
In summary, to calculate $\Delta E^{\rm epi}(a_s,\widehat{G})$ of
Eq.~(\ref{eq:simpleepi}) we will use Eq.~(\ref{eq:simple_strainE})
to obtain it from LDA for a few substrate lattice parameters $a_s$ and
along selected symmetry directions $\widehat{G}$. 
We will also need to obtain the harmonic elastic constants
$C_{11}$, $C_{12}$ and $C_{44}$. 
The calculated $\Delta E^{\rm epi}(a_s,\widehat{G})$ 
results are then fitted by the general Eqs.~(\ref{eq:qdef}),
(\ref{eq:gamma_general}) and (\ref{eq:q_general}).

\subsection{Anharmonic epitaxial strain of thin films of pure 
elements: LDA results}
\label{sec:qs}

\begin{figure*}[th]
\epsfxsize=5.5in
\centerline{\epsffile{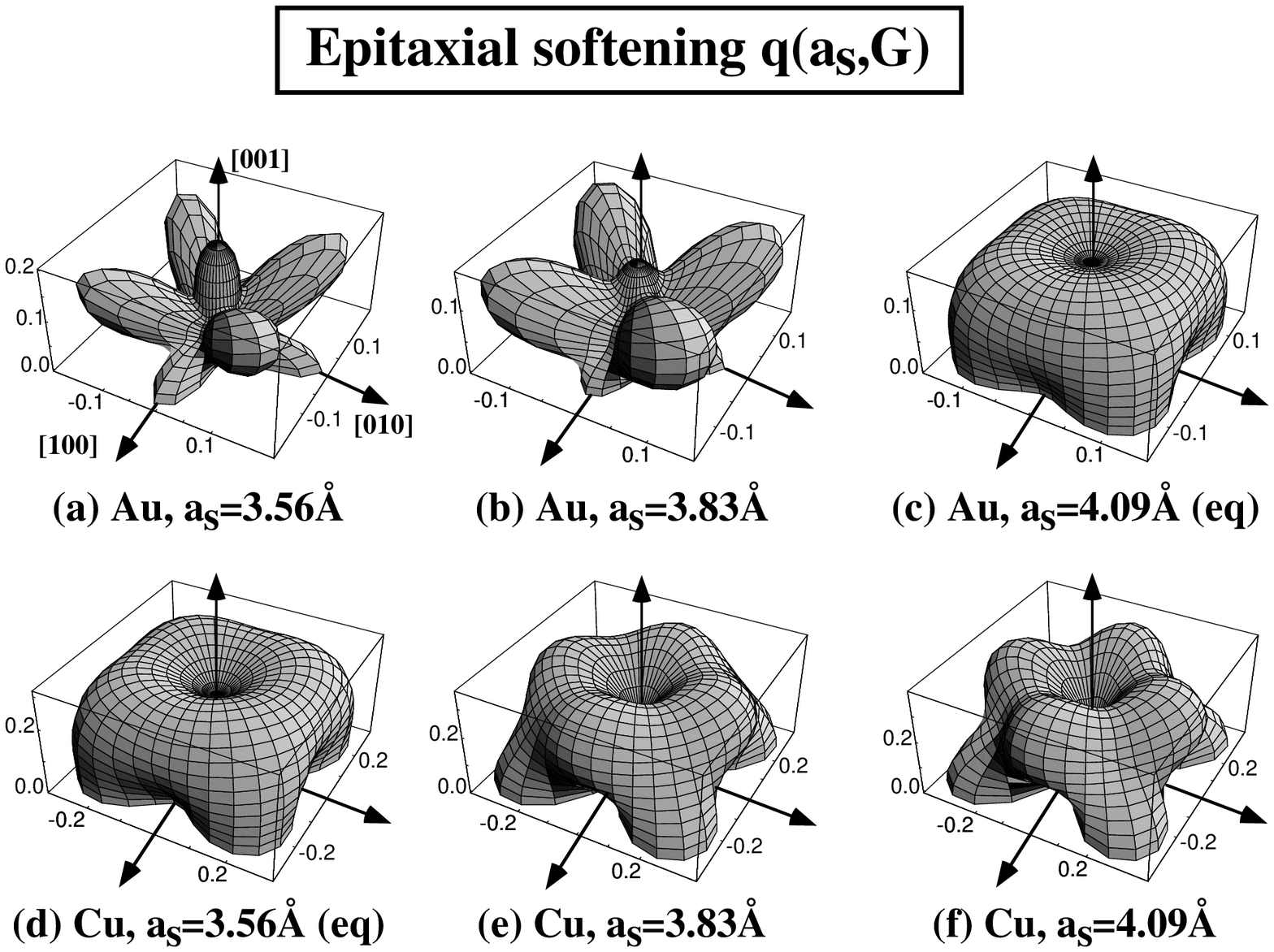}}
\vskip 5mm
\caption{Epitaxial softening function $q(a_s,\widehat{G})$ for
(a)--(c) Cu and (d)--(e) Au, at different values of the substrate
lattice constant $a_s$.}
\label{fig:qsexy}
\end{figure*}

We have calculated the epitaxial strain energy 
$\Delta E^{\rm epi} (a_s,\widehat{G})$
for Cu, Ni, Ag and Au along six principle directions
$\langle 001 \rangle$, $\langle 111 \rangle$, 
$\langle 110 \rangle$, $\langle 113 \rangle$, 
$\langle 201 \rangle$ and $\langle 221 \rangle$.
The local-density approximation\cite{DF} (LDA), as implemented
by the linearized augmented plane wave (LAPW) method\cite{LAPW}, was
used to obtain the total energies in Eqs.~(\ref{eq:simple_strainE}) and
(\ref{eq:qdef}).
$q(a_s,\widehat{G})$ was calculated from Eq.~(\ref{eq:qdef})
and fitted with the functional form
Eqs.~(\ref{eq:gamma_general})--(\ref{eq:q_general}).
The angular momentum cutoff in Eq.~(\ref{eq:gamma_general}) was set to
$l_{\rm max}=10$,
leaving five independent coefficients for each value of the substrate
lattice parameter $a_s$. This choice allows reproduction of the LDA
values with a maximum error of $0.04$. The calculations
have been done for biaxial compression ($a_s < a_{\rm eq}$) of Au and
Ag, for biaxial expansion ($a_s > a_{\rm eq}$) of Ni, and for both
biaxial expansion and compression of Cu. The expansion
coefficiets $b_l (a_s)$, entering Eqs.~(\ref{eq:gamma_general}), are
shown in Fig.~\ref{fig:bl}.  At the equilibrium lattice constant
$a_{\rm eq}$ (vertical arrows in Fig.~\ref{fig:bl}), where the
harmonic formula Eq.~(\ref{eq:gamma_harm}) is
exact, all $b_l$ are exactly zero. As $a_s$ deviates from 
$a_{\rm eq}$, they change rapidly indicating the importance of
anharmonic effects. In Cu and Ni for $a_s > a_{\rm eq}$, $l=6$ 
term is as important as $l=0$ and $l=4$ terms, contributions
from $l \ge 8$ being an order of magnitude smaller. In Au
for $a_s < a_{\rm eq}$, $b_0(a_s)$ and $b_8(a_s)$ are the dominating
terms, while the behavior of Ag is mainly determined by $b_0(a_s)$
and $b_6(a_s)$. Thus, in spite of broad similarities
between the studied elements, they exhibit some interesting
differences.

Figure~\ref{fig:qs} shows the calculated LDA epitaxial softening
functions $q(a_s,\widehat{G})$ of Eq.~(\ref{eq:qdef}) for Cu, Ni, Ag
and Au. There are important qualitative and quantitative differences
between $q_{\rm harm} (\widehat{G})$ given by the harmonic elasticity
Eq.~(\ref{eq:qharm}), and the anharmonic $q(a_s,\widehat{G})$
calculated from the LDA.
First, all $q(a_s,\widehat{G})$ depend on the
substrate lattice constant $a_s$, while the harmonic  $q_{\rm harm}
(\widehat{G})$ are independent of $a_s$. Figure~\ref{fig:qsexy}
shows the directional dependence of $q(a_s,\widehat{G})$ for Cu and Au
at a few values of $a_s$: equilibrium lattice parameter of Cu (3.56~\AA),
equilibrium lattice parameter of Au (4.04~\AA), and halfway between
them (3.83~\AA). 
By construction, $q$ at $a_s=a_{\rm eq}$ is given by the
harmonic form Eqs.~(\ref{eq:qharm})--(\ref{eq:gamma_harm}), shown
for fcc Au in Fig.~\ref{fig:qsexy}(c) and fcc Cu in
Fig.~\ref{fig:qsexy}(d). Epitaxial deformation
of Au with $a_s < a_{\rm eq}$ makes the lobes along 
$\langle 111 \rangle$ much more
pronounced than in the harmonic case. Furthermore, $q$ for Au develops
additional lobes along $\langle 001 \rangle$, which in the 
harmonic approximation is
the softest direction. In contrast, $q$ of Cu under biaxial
expansion exhibits pronounced deepening of the $\langle 001 \rangle$
minima, but develops maxima along $\langle 110 \rangle$.

Second, in the harmonic elasticity theory of Eq.~(\ref{eq:qharm})
if $\langle 001 \rangle$ is the softest direction 
(smallest $q_{\rm harm}$), then $\langle 111 \rangle$ {\it
must\/} be the hardest direction, and vice versa. Figure~\ref{fig:qs}
shows that this order does not hold for large deformations: 
the hardest direction in Ni and Cu for $a_s \gg a_{\rm eq}$ 
is $\langle 110 \rangle$, while the hardest directions 
in Ag and Au for $a_s \ll a_{\rm eq}$ are $\langle 111 \rangle$ and 
$\langle 001 \rangle$, both $\langle 110 \rangle$ and 
$\langle 201 \rangle$ being much softer than the former.

\begin{figure*}[th]
\epsfxsize=5.5in
\centerline{\epsffile{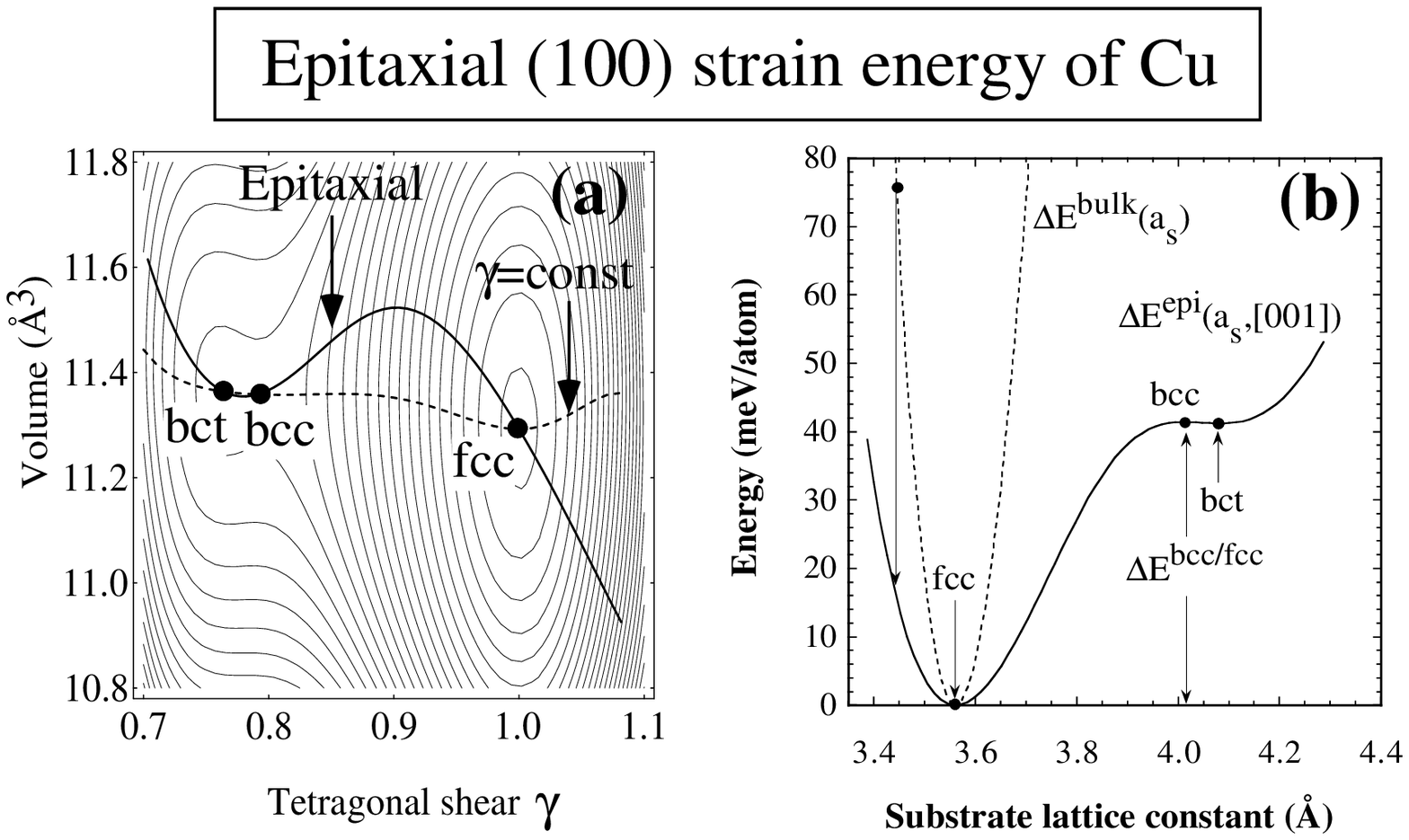}}
\vskip 5mm
\caption{Contour plot of the two-dimensional energy surface
$E(\gamma,V)$ for Cu. The continuous line shows the
epitaxial path determined by Eq.~(\protect\ref{eq:baines1}), while the
dashed line is the relation $V=V(\gamma)$ obtained by minimizing
$E(\gamma,V)$ with respect to the volume $V$ at a constant $\gamma$.
The right panel shows the epitaxial strain energy as a function
of the substrate lattice constant in comparison with the
(much larger) bulk deformation energy $\Delta E^{\rm bulk}(a_s)$.}
\label{fig:E(c/a)}
\end{figure*}

Third, Fig.~\ref{fig:qs} shows that $q(a_s,\widehat{G})$ of different
directions cross for substrate/film lattice mismatch
$ 2 |a_s-a_{\rm eq}| / |a_s+a_{\rm eq}| < 4\%$. For example, while
$\langle 001 \rangle$ is the softest direction near $a_{\rm eq}$ 
and stays such upon biaxial expansion (Cu, Ni), it is one of the
hardest in biaxially compressed metals (Ag, Au, Cu) where 
$\langle 201 \rangle$ is the softest direction.
Similarly, $\langle 111 \rangle$ is the hardest direction near the
equilibrium and for $a_s \ll a_{\rm eq}$, but it becomes softer than
$\langle 110 \rangle$ and $\langle 201 \rangle$ 
in biaxially expanded Cu and Ni.
Thus, there is a qualitative breakdown of the harmonic theory for
strains of $4\%$, and presumably quantitative errors for even smaller
strains.

We also note 
similarities in the elastic behavior of these materials. Under
expansion, both Cu and Ni exhibit strong softening of 
$q(a_s,[001])$ and somewhat weaker softening of $q(a_s,[111])$, 
while $q(a_s,[110])$ becomes the 
elastically hardest direction. This order is reversed under
biaxial compression of Ag, Au and Cu: $q$'s for $\langle 001 \rangle$
and $\langle 111 \rangle$ 
harden, but the $\langle 110 \rangle$ and $\langle 201 \rangle$ 
directions soften.

\subsection{Discussion of anomalous softening of $q(a_s,\widehat{G})$
in terms of fcc/bcc energy differences}
\label{sec:fcc/bcc}

The anomalous softening of $q(a_s,[001])$ in Ni and Cu for 
$a_s > a_{\rm eq}$ reflects a small fcc/bcc energy difference for
these materials. This can be seen by considering three energy surfaces
that deform fcc into bcc:

(i) $E(\gamma,V)$: The most general surface is the total energy as a
function of the tetragonal shear $\gamma$ and volume $V$, shown 
as contour in
Fig.~\ref{fig:E(c/a)}(a) for Cu. The tetragonal shear along 
$\langle 001 \rangle$ is defined by:
\begin{eqnarray}
\varepsilon_{ij} = \left(
\begin{array}{ccc}
\gamma & 0 & 0\\
0 & \gamma^{-\frac{1}{2}} & 0\\
0 & 0 & \gamma^{-\frac{1}{2}}\\
\end{array}
\right),
\end{eqnarray}
where $c/a = \sqrt{2} \gamma^\frac{3}{2}$.
$E(\gamma,V)$ has (at least) three extremal points, denoted in
Fig.~\ref{fig:E(c/a)}(a) as solid dots: one corresponding 
to the fcc state, one to the bcc state and one to the bct state.
These states obey the extremal conditions of vanishing derivatives:
\begin{equation}
\label{eq:baines3}
\frac{\partial}{\partial \gamma} E(\gamma, V) =
\frac{\partial}{\partial V}    E(\gamma, V) = 0.
\end{equation}
Figure~\ref{fig:E(c/a)}(a) shows that for Cu fcc and bct are locally
stable minima with respect to $\gamma$ and $V$, while bcc is a saddle
point (maximum with respect to $\gamma$ and minimum with respect to
$V$).\cite{kraft93,alippi}

(ii) Bain path $E(\gamma)$: A more specific function 
$E(\gamma) \equiv \left. E(\gamma,V) \right|_{V={\rm const}}$ 
is defined by the tetragonal Bain
path,\cite{bain24} connecting fcc and bcc structures. The Bain path is
obtained by changing the $c/a$ ratio while keeping $V=ca^2$ constant.
When $c/a=1$ the lattice type is bcc and when
$c/a=\sqrt{2}$ it is fcc. The energy as a function of $\gamma$ must have
extremal points at both $\gamma$ values corresponding to the cubic
symmetry fcc ($\gamma_{\rm fcc} = 1$) and 
bcc ($\gamma_{\rm bcc} = 2^{-\frac{1}{3}}$) states,
as well as at least another bct point $\gamma_{\rm bct}$ with a zero
derivative $E'(\gamma)=0$.\cite{sob97,explainBain}
Usually,\cite{kraft93,alippi,sob97,craievich94,craievich97,ptw}
for fcc stable elements the bcc lattice is unstable 
[i.e., $E(\gamma)$ has a local maximum at $\gamma_{\rm bcc}$] and
the bct state (a local minimum) occurs for $\gamma_{\rm bct} <
\gamma_{\rm bcc}$.

(iii) Epitaxial Bain path $E[c_{\rm eq}(a_s)]$: This deformation path
is obtained by scanning $c$ while $a_s$ is kept fixed, which
corresponds to epitaxial growth on a (001) substrate with lattice
parameter $a_s$. $c$ is determined from the total energy minimization
at a fixed $a_s$: 
\begin{equation}
\label{eq:baines1}
\frac{d}{dc} E^{\rm tot} (\gamma,V)
= \left( \frac{2}{3} \gamma^{-\frac{1}{3}}
	\frac{\partial}{\partial \gamma} +
a^2_s \frac{\partial}{\partial V} \right) E(\gamma,V) = 0.
\end{equation}
Eq.(\ref{eq:baines1}) defines {\it the epitaxial path\/}
$V(\gamma)$, shown as a continuous line in Fig.~\ref{fig:E(c/a)}(a).
Since $c/a_s = \sqrt{2} \gamma^\frac{3}{2}$ and 
$V=c a_s^2/4$, this path implicitly
relates the out-of-plane dimension $c$
to the substrate lattice constant $a_s$, much 
like Eq.~(\ref{eq:c/a}) does in the harmonic case.
As noted in Ref.~\onlinecite{alippi},
the epitaxial path crosses all extremal points
of $E(\gamma,V)$ because Eq.~(\ref{eq:baines1}) is satisfied
where conditions Eq.~(\ref{eq:baines3}) hold.
Therefore, if we parametrize the epitaxial strain energy along this
path as a function of $a_s$, it
has a global minimum corresponding to fcc, a locally stable
minimum corresponding to bct and a maximum at the bcc state,
see Fig.~\ref{fig:E(c/a)}(b). 
We see that as $a_s$ increases from the equilibrium 
fcc value, Cu sequentially passes through the bcc and bct states
where the strain energy $\Delta E^{\rm epi}(a_s,[001])$ 
is equal to the fcc/bcc and fcc/bct structural
energy differences. When these energy differences
are much smaller than the characteristic
values of the bulk deformation energies $\Delta E^{\rm bulk} (a_s)$
[see Fig.~\ref{fig:E(c/a)}(b)], then $q(a_s,[001])$ is anomalously soft
[since $q(a_s,[001])= \Delta E^{\rm bcc/fcc} / \Delta E^{\rm bulk}_{\rm
fcc} (a_s)$ for $a_s = (2V_{\rm bcc})^\frac{1}{3}$].

In summary, the softness of $q(a_s,[001])$ for $a_s > a_{\rm eq}$
is a reflection of the geometric properties of the
$\langle 001 \rangle$ epitaxial
deformation path (connection between {\it cubic symmetry\/} fcc and
bcc structures), and a small fcc/bcc energy difference,
$\Delta E^{\rm fcc/bcc} \ll \Delta E^{\rm bulk} (a_s)$.
It is important that the fcc and bcc points correspond to
lattices with cubic symmetry, since it ensures that the
energy surface has extremal points there. In zincblende GaP and
InP,\cite{shwei}
epitaxial $\langle 001 \rangle$ path has only one point of
cubic symmetry ($c/a=\sqrt{2}$, corresponding to undistorted fcc),
and therefore the energy surface $E(\gamma,V)$ is not required
to possess additional extremal points. As a consequence, $\Delta
E^{\rm epi}(a_s,[001])$ is a monotonously increasing function
of $a_s$, and $q(a_s,[001])$ does not soften with increasing $a_s$.

The described mechanism also accounts for the
softening of $q(a_s,[111])$ for $a_s>a_{\rm eq}$ in Cu and Ni under 
biaxial $\langle 111 \rangle$ expansion, since this deformation  
takes fcc ($c/a=\sqrt{6}$) into bcc ($c/a=\sqrt{6}/4$), albeit at a
much larger strain. However, we have not found any simple structure
corresponding to the compressive $\langle 201 \rangle$ strain which
could explain the softening of  
$q(a_s < a_{\rm eq},[201])$ in Ag, Au and Cu. The latter seems to be
caused by relatively loose packing of atoms within the (201) planes,
imposing small energy penalty on decreasing the interatomic distances.
Indeed, the nearest-neighbor distance in (201) plane is $a_s$,
compared to $a_s/\sqrt{2}$ in (111) or (001) planes with high values
of $q(a_s,\widehat{G})$ for $a_s < a_{\rm eq}$.

\section{Stability of superlattices and alloys}
\label{sec:SLs}

\subsection{Constituent strain of superlattices}
\label{sec:Hinfinite}

\begin{figure}
\epsfxsize=2.8in
\centerline{\epsffile{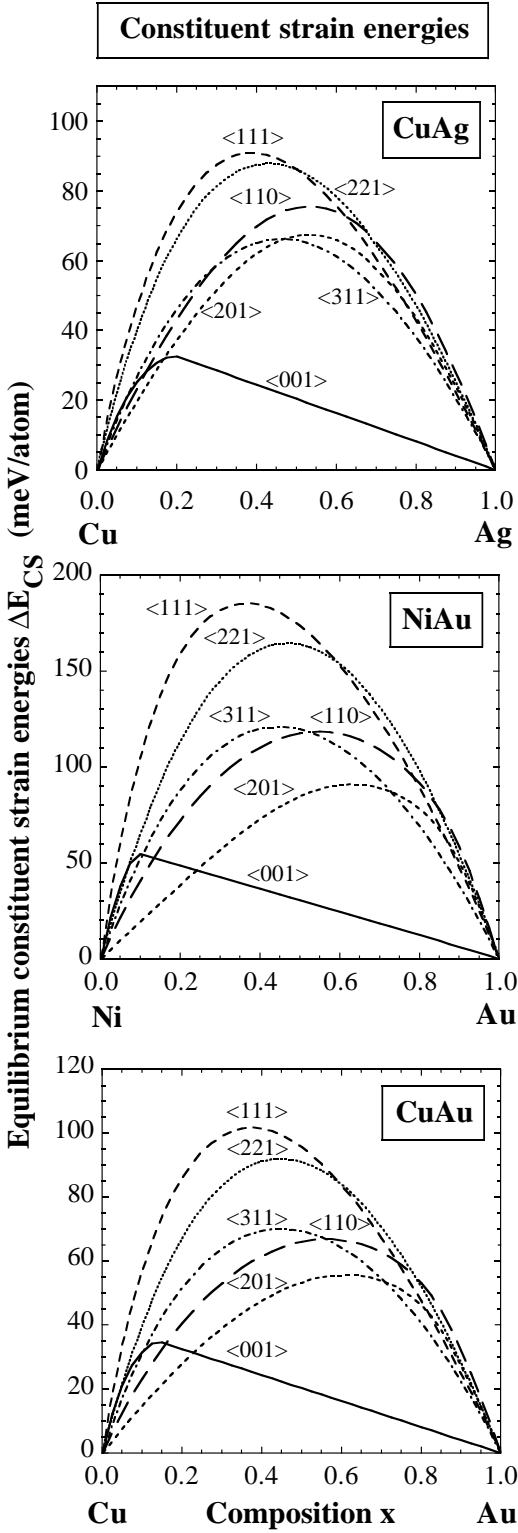}}
\vskip 5mm
\caption{Equilibrium constituent strain energies for Cu-Au, Ni-Au and
Cu-Ag. Ag-Au system is size matched and $\Delta E^{\rm eq}_{\rm CS}=0$.}
\label{fig:Ecs}
\end{figure}

\begin{figure}
\epsfxsize=2.8in
\centerline{\epsffile{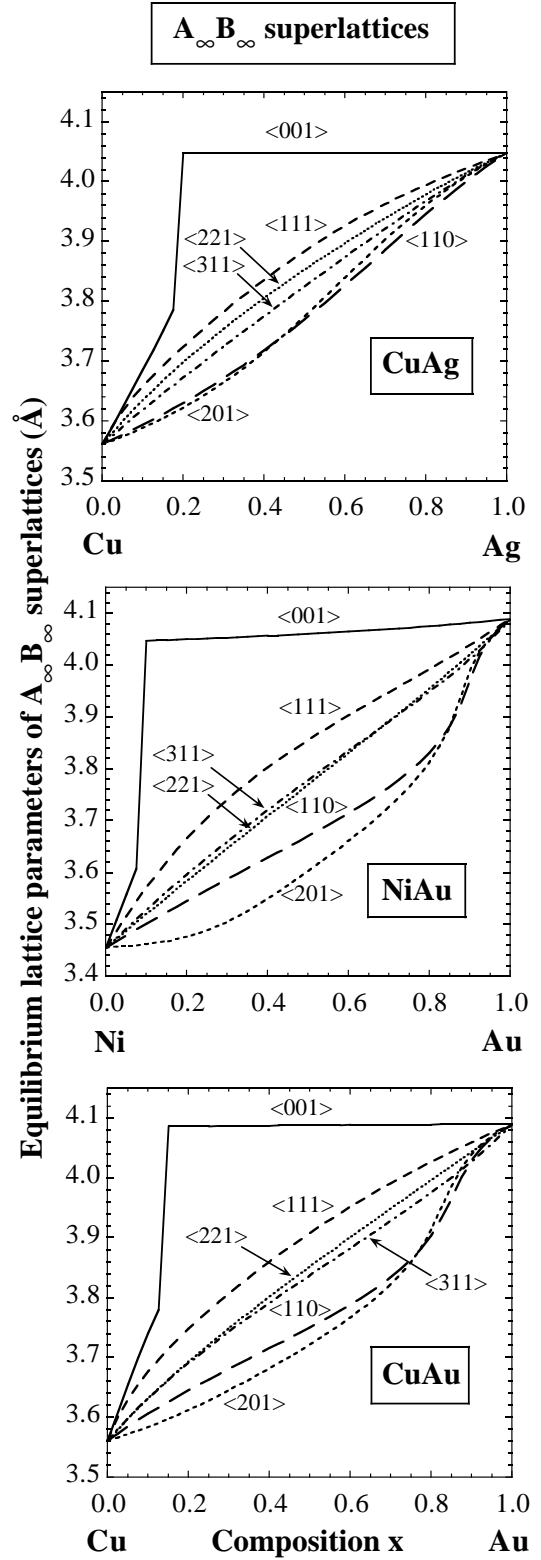}}
\vskip 5mm
\caption{Equilibrium lattice parameter of infinite Cu-Au, Cu-Ag and
Ni-Au superlattices vs composition.}
\label{fig:alat_SL}
\end{figure}

The bulk formation enthalpy of superlattices
[Eq.~(\ref{eq:Einterface})]
is expressed as a sum of the interfacial energy 
$I(n,\widehat{G})$ and constituent strain energy 
$\Delta E_{\rm CS}^{\rm eq} (x,\widehat{G})$.
As given by Eq.~(\ref{eq:ECSdef}), the latter is a
weighted average of the epitaxial strain energies of coherently
strained constituents, minimized with respect to the
common in-plane lattice vectors ${\bf a}_1$ and ${\bf a}_2$. 
For the high symmetry directions 
$\langle 001 \rangle$ and $\langle 111 \rangle$,
these vectors are related by symmetry operations of the superlattice,
so that  ${\bf a}_1$ and ${\bf a}_2$  are 
proportional to the ideal fcc unit vectors ${\bf a}_1^0$ and 
${\bf a}_2^0$ via Eq.~(\ref{eq:scaling}). Then 
$\Delta E_{\rm CS}^{\rm eq} (x,\widehat{G})$
can be calculated by minimizing the following expression
with respect to the superlattice parameter $a_{\rm SL}$:
\begin{eqnarray}
\nonumber
\Delta E_{\rm CS}^{\rm eq} (x, \widehat{G}) = 
\min_{a_{\rm SL}} \bigl[ (1-x) 
	\Delta E^{\rm epi}_A (a_{\rm SL},\widehat{G})\\
\label{eq:ECShighsymm}
 + x \Delta E^{\rm epi}_B (a_{\rm SL},\widehat{G}) \bigr].
\end{eqnarray}
For lower symmetry directions $\widehat{G}$, the in-plane unit vectors
${\bf a}_1$ and ${\bf a}_2$ may relax differently, and the angle
$\cos \gamma = {\bf a}_1 \cdot {\bf a}_2 / |{\bf a}_1| |{\bf a}_2|$
is also free to vary. For instance, in 
$\langle 110 \rangle$ superlattices, the vectors
${\bf a}_1$ and ${\bf a}_2$ are not related by symmetry,
and therefore may scale differently, i.e., in ideal fcc 
$|{\bf a}_1^0| / |{\bf a}_2^0| = \sqrt{2}$ but in the superlattice
generally $|{\bf a}_1| / |{\bf a}_2| \ne \sqrt{2}$.
Equation~(\ref{eq:ECShighsymm}) is much simpler than the general
Eq.~(\ref{eq:ECSdef}) requiring minimization with respect to {\it
three\/} degrees of freedom: lengths $|{\bf a}_1|$, $|{\bf a}_2|$, and
the angle  $\alpha = \widehat{({\bf a}_1, {\bf a}_2)}$. 
In the present work we adopt Eq.~(\ref{eq:ECShighsymm}) even for low
symmetry directions, using the calculated $\Delta E^{\rm
epi}_A(a_s,\widehat{G})$ from Sec.~\ref{sec:qs}.

$\underline{\Delta E_{\rm CS}^{\rm eq}}$: 
Figure~\ref{fig:Ecs} shows the equilibrium constituent 
strain energies $\Delta E_{\rm CS}^{\rm eq}(x,\widehat{G})$
for the size-mismatched Cu-Ag, Ni-Au and Cu-Au systems.
They are determined from  Eq.~(\ref{eq:ECShighsymm}),
using only the epitaxial strain energies $\Delta E^{\rm epi}_{A,B}$
of the constituents. There are obvious similarities in 
$\Delta E_{\rm CS}^{\rm eq}(x,\widehat{G})$ for the three noble
metal systems. $\langle 201 \rangle$ superlattices have the lowest
constituent strain energy below $x \approx 0.2$,
after that $\langle 001 \rangle$ becomes the softest direction. 
$\langle 111 \rangle$ is the hardest 
direction over a wide composition range, except close to $x=1$
where $\langle 110 \rangle$ is slightly harder. 

This behavior can be explained by
the properties of the epitaxial softening function
$q(a_s,\widehat{G})$, discussed in Sec.~\ref{sec:qs}.
For example, consider Cu-Au from Fig.~\ref{fig:Ecs}. Upon
biaxial compression of Au (corresponding to 
$x < 0.5$), $q(a_s,[111])$ increases rapidly (see
Fig.~\ref{fig:qs}), increasing the elastic strain energy
and making this an elastically hard direction. In contrast,
$q(a_s,[201])$ for Au decreases with biaxial compression, and 
at $x <0.2$ there is small energetic penalty for deforming Cu and Au
to a common in-plane lattice constant. Increase of $q(a_s,[110])$ 
for Cu with $a_s$ eventually causes this to be the hardest direction
in Au-rich Cu-Au superlattices.

$\underline{a_{\rm SL} (x)}$: Figure~\ref{fig:alat_SL} shows the
equilibrium in-plane lattice constant $a_{\rm SL}(x,\widehat{G})$
that minimizes the constituent strain.
These are also the equilibrium lattice parameters for infinite period
superlattices.
The lattice parameters $a_{\rm SL} (x,\widehat{G})$ show large
deviations from Vegard's law, with the behavior of $a_{\rm SL}
(x,[001])$ being particularly anomalous.
The very unusual composition dependence of the superlattice parameter
for $\langle 001 \rangle$ deserves a closer scrutiny:
At $x \approx 0.2$ the superlattice parameter changes discontinuously
to the lattice parameter of the larger constituent. The constituent
strain energy abruptly changes slope 
and settles down to a strictly linear composition
dependence. Furthermore, $\Delta E_{\rm CS}^{\rm eq} (x,[001])$ is
very small in comparison with $\Delta E_{\rm CS}^{\rm eq}$ for other
directions. These anomalies are direct consequences 
of the soft $q(a_s,[001])$ for biaxially expanded Cu and Ni, 
which in turn is a consequence of the
small fcc/bcc and fcc/bct energy differences for these metals
(Sec.~\ref{sec:qs}). Indeed, for a sufficiently Au-rich system
$\Delta E^{\rm epi}_{\rm Cu}$ is smaller than 
$\Delta E^{\rm epi}_{\rm Au}$ favoring a
superlattice constant close to the equilibrium lattice parameter of
Au. This large lattice parameter
happens to fall on the flat region of the strain energy
curve around the bcc and bct states of biaxially expanded Cu (see
Fig.~\ref{fig:E(c/a)}), where a local bct minimum exists in
the function on the right-hand side of Eq.~(\ref{eq:ECShighsymm}),
shifting downward in energy with increasing $x$. At some critical
value of the composition, the minimum around $a_{\rm Au}$
becomes deeper than the minimum close to $a_{\rm Cu}$, which causes a
discontinuous jump in $a_{\rm SL}$. Loosely speaking, Cu deforms
all the way into the bct structure and Au does not deform at all. That
also explains the linear decrease of $\Delta E_{\rm CS}(x,[001])$
after the discontinuity, since $\Delta E^{\rm epi}_{\rm Au} = 0$ and
$\Delta E^{\rm epi}_{\rm Cu} = {\rm const}$ in Eq.~(\ref{eq:ECShighsymm}).

In conclusion, we summarize the prerequisites for low 
elastic strain energy of infinite superlattices:

(i) One of the components should exhibit a particularly soft elastic
direction under biaxial strain, e.g., $\langle 001 \rangle$ in Cu upon
epitaxial expansion and $\langle 201 \rangle$ in Au upon biaxial
compression.

(ii) The lattice mismatch between the constituents should be large
enough to explore the regions of anomalous softness. 

We stress that the unusual behavior
shown in Figs.~\ref{fig:alat_SL} and \ref{fig:Ecs}
(crossing of different directions, discontinuities, different
skewnesses of $\Delta E_{\rm CS}^{\rm eq}
(x,\widehat{G})$ curves) are caused by the
anharmonic $q(a_s,\widehat{G})$, and cannot be obtained
within the harmonic theory with lattice parameter
independent $q_{\rm harm} (\widehat{G})$.\cite{dblaks92}

\subsection{Describing chemical interactions via the mixed-space
cluster expansion} 
\label{sec:CE}

The energy of a bulk alloy $\Delta H_{\rm mix}^{\rm bulk} (x)$
of Eq.~(\ref{eq:Hmix}), and of an epitaxial alloy 
$\Delta H_{\rm mix}^{\rm epi} (x)$ of Eq.~(\ref{eq:epitaxialHmix})
cannot be computed directly from LDA since configurationally
random structures are involved. The approximate approach is
either large supercells or a first-principles mixed-space cluster 
expansion.\cite{dblaks92,NATO} In the latter approach, a spin variable
$S_i$ is assigned to each lattice site ${\bf R}_i$ which takes a value
$+1$ if the site is occupied by an atom of type $A$, or $-1$ if the
site is occupied by an atom of type $B$. The formation enthalpy of an
arbitrary structure $\sigma$ is expressed in the following form:
\begin{eqnarray}
\nonumber
\Delta H_{\rm CE} (\sigma) = \sum_{\bf k} J_{\rm pair} ({\bf k})
\, \left| S({\bf k},\sigma) \right|^2 \\
\label{eq:recipCE}
+ \sum_f^{\rm MB} D_f J_f \overline{\Pi}_f (\sigma) 
+ \Delta E_{\rm CS} (\sigma).
\end{eqnarray}
where $J({\bf k})$ is the Fourier transform of real-space pair
interactions and $S({\bf k},\sigma)$ is the structure factor,
\begin{eqnarray}
\label{eq:jofk}
J_{\rm pair} ({\bf k}) = \sum_j J_{\rm pair} ({\bf R}_i-{\bf R}_j)
e^{-i{\bf kR}_j},\\
\label{eq:strfac}
S({\bf k}, \sigma) = \sum_j S_j e^{-i{\bf kR}_j}.
\end{eqnarray}
The second sum in Eq.~(\ref{eq:recipCE}) runs over symmetry
inequivalent clusters constituted by three or more lattice
sites. $D_f$ is the number of equivalent clusters per lattice site,
and $\overline{\Pi}_f (\sigma)$ are structure-dependent 
geometrical coefficients (simple lattice averages of the cluster
spin products).
The last term in Eq.~(\ref{eq:recipCE}) is the constituent strain
energy  $\Delta E_{CS} (\sigma)$ of the structure $\sigma$. It
is designed to reproduce the elastic strain energy of coherent
long-period superlattices\cite{dblaks92} which are calculated
directly from the LDA (see Sec.~\ref{sec:Hinfinite}):
\begin{eqnarray}
\label{eq:ECSsigma}
\Delta E_{\rm CS} (\sigma) = \sum_{\bf k} J_{\rm CS} (x,\widehat{k})
\left| S({\bf k},\sigma) \right|^2,\\
\label{eq:Jcsdef}
J_{\rm CS} (x,\widehat{k}) = \frac{\Delta E^{\rm eq}_{\rm
CS}(x,\widehat{k})} {4x(1-x)}.
\end{eqnarray}
The equilibrium constituent strain energies 
$\Delta E^{\rm eq}_{\rm CS}(x,\widehat{k})$
have been deduced from the directly calculated
$\Delta E^{\rm epi} (a_{\rm SL},\widehat{G})$ minimizing
Eq.~(\ref{eq:ECShighsymm}) with respect to the common
in-plane lattice constant $a_{\rm SL}$.
They are fitted by series of Kubic harmonics
with composition dependent coefficients $c_l(x)$:
\begin{equation}
\label{eq:Ecs_expansion}
\Delta E_{\rm CS} (x,\widehat{G}) = \sum_{l=0}^{l_{\rm max}}
c_l(x) \, K_l (\widehat{G}),
\end{equation}
which are used to evaluate $\Delta E_{\rm CS} (x,\widehat{G})$
for any direction $\widehat{G}$.
Structure factors $S({\bf k},\sigma)$ in the long-period superlattice
limit are nonzero only for ${\bf k} \rightarrow 0$, where
$J_{\rm CS} (x,\widehat{k})$ is a nonanalytic function of ${\bf k}$,
reflecting the directional dependence of the constituent strain
energy.

The {\it effective cluster interactions\/} $J_f$ and $J_{\rm
pair} ({\bf k})$ are determined by fitting Eq.~(\ref{eq:recipCE})
to a large number (30 to 40) fully relaxed first-principles
LDA formation enthalpies of simple ordered structures.
Most of these ordered structures are short-period superlattices along
$\langle 001 \rangle$, $\langle 111 \rangle$, 
$\langle 110 \rangle$, $\langle 201 \rangle$ and 
$\langle 113 \rangle$.\cite{paperI}
The calculations of $T=0$ total energies employ the full-potential
linearized augmented plane wave method\cite{lapw} (FLAPW) and local
density approximation (LDA) for the electronic
exchange and correlation. The total energy is minimized with respect
to all structural degress of freedom, i.e. both the atomic positions
and cell-external coordinates are fully relaxed. Complete discussion of
the LDA calculations and cluster expansions for Ag-Au, Cu-Ag, Cu-Au
and Ni-Au can found in Ref.~\onlinecite{paperI}.

\subsection{Stability of finite period metal superlattices}
\label{sec:finiteSL}

\begin{figure}
\epsfxsize=3in
\centerline{\epsffile{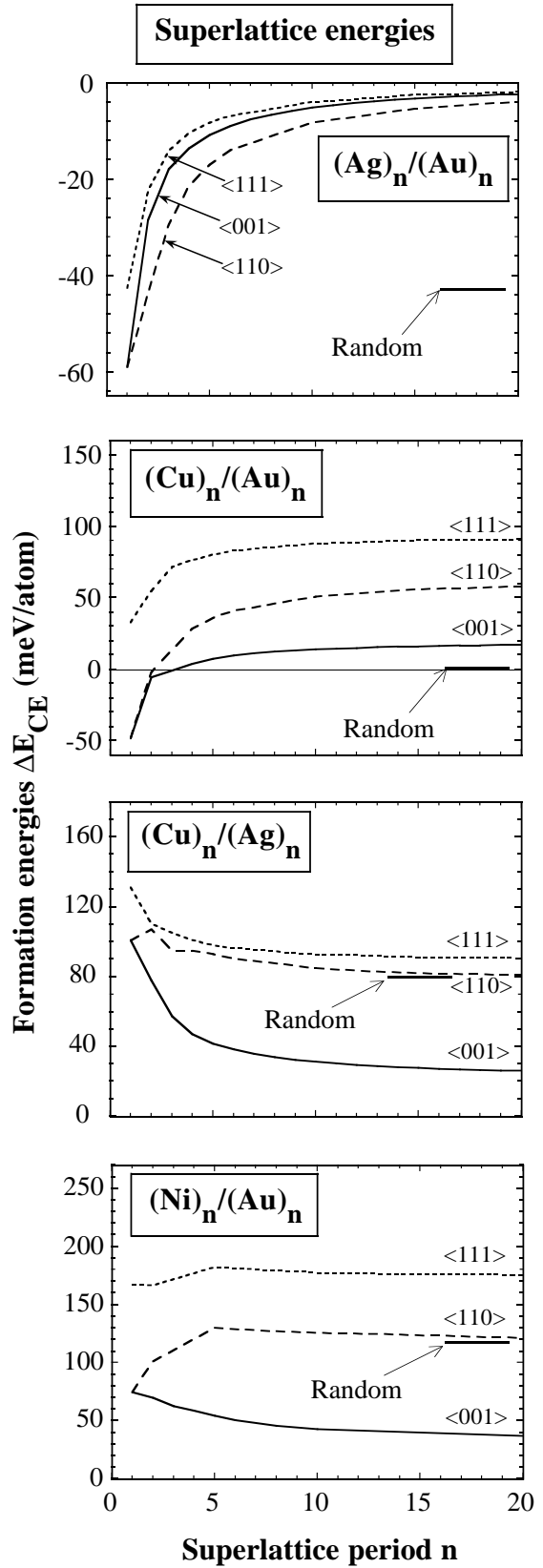}}
\vskip 5mm
\caption{Superlattice energies for Cu-Au, Cu-Ag, Ni-Au and Ag-Au.}
\label{fig:SL}
\end{figure}

\begin{figure}
\epsfxsize=3in
\centerline{\epsffile{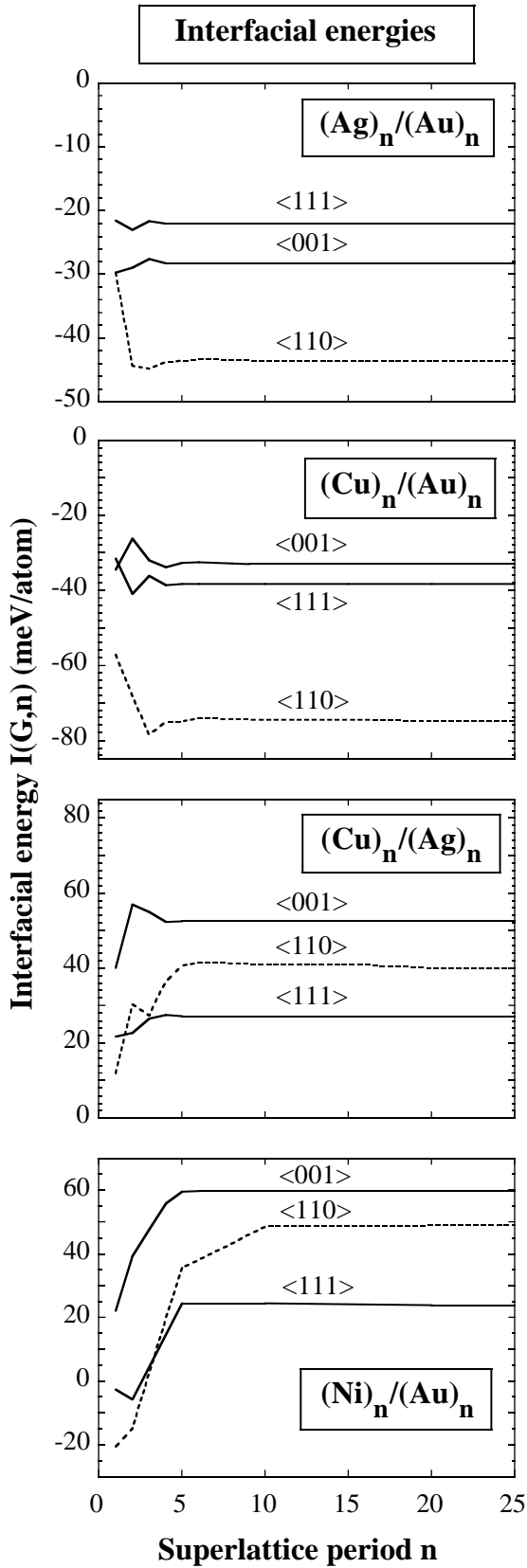}}
\vskip 5mm
\caption{Interfacial energies of Cu-Au, Cu-Ag, Ni-Au and Ag-Au.}
\label{fig:interf}
\end{figure}

Having obtained all ingredients of $\Delta H_{\rm CE}(\sigma)$
[Eq.~(\ref{eq:recipCE})] from LDA calculations on small unit cell
structures, we can use this equation to predict the energy of {\it
any\/} configuration $\sigma$, in particular superlattices.
Figure~\ref{fig:SL} shows the bulk formation energies
of $(A)_n/(B)_n$ superlattices for the studied noble metal
systems. The interfacial energies $I(n,\widehat{G})$, extracted from 
$\Delta H_{SL} (n,\widehat{G})$ according to Eq.~(\ref{eq:Einterface}),
are shown in Fig.~\ref{fig:interf}. Several interesting
observations can be made from these figures:

(i) $I(n,\widehat{G})$ are approximately constant after $n > 5$.

(ii) For ordering systems (Cu-Au and Ag-Au), the interfacial
energies are negative (see Fig.~\ref{fig:interf}). Negative
interfacial energies are the cause for the upward slope of 
$\Delta H_{\rm SL}(n,\widehat{G})$ curves in 
Fig.~\ref{fig:SL}. This indicates a chemical preference for having 
unlike atoms at the interface and a tendency to form ordered
structures at $T=0$. For instance, $L1_0$, the observed ground state 
of CuAu, is a monolayer (Cu)/(Au) superlattice along $\langle 001
\rangle$. The order of $\Delta H_{\rm SL} (n,\widehat{G})$ is lowest 
$\langle 001 \rangle$ and highest $\langle 111 \rangle$ for Cu-Au, and
lowest $\langle 110 \rangle$ and highest $\langle 111 \rangle$ for
Ag-Au superlattices.

(iii) For the phase separating Cu-Ag, all
interfacial energies are positive.
$\Delta H_{\rm SL}(n,\widehat{G})$ decreases with $n$ for all 
directions and reflect the tendency to coherent phase separation
over ordered superlattice formation. Interfaces between Cu and Ag are
energetically very costly.
The order of $\Delta H_{\rm SL} (n,\widehat{G})$ is again lowest 
$\langle 001 \rangle$ and highest $\langle 111 \rangle$.

(iv) Ni-Au has the most interesting behavior of 
$\Delta H_{\rm SL}(n,\widehat{G})$ and $I(n,\widehat{G})$.
It exibits phase-separating type 
$\Delta H_{\rm SL}(n,[001])$ (decreasing with $n$),
ordering type $\Delta H_{\rm SL}(n,[110])$ (increasing with $n$),
and a nearly constant  $\Delta H_{\rm SL}(n,[111])$. Does this mean
that interfaces in some directions are energetically favorable, while
in other directions they are energetically costly?
The answer is: No. In Ni-Au, just like in Cu-Ag,
all isolated interfaces have positive formation energies. However,
{\it the interaction between the interfaces along $\langle 110
\rangle$ is strongly attractive in Ni-Au\/}, and leads to a net
chemical energy gain for some short-period superlattices. Indeed,
Fig.~\ref{fig:interf} shows that all interfacial energies
of Ni-Au are positive in the limit $n \rightarrow \infty$ (when there
is no interaction between the interfaces), but
decrease for short periods and are negative for 
$\langle 110 \rangle$ $n \leq 3$
superlattices. As we show in Ref.~\onlinecite{paperSRO}
the competition between the constituent strain energy, interfacial
energy $I(n \rightarrow \infty,\widehat{G})$ and ordering-type
interaction between the interfaces leads to unusual
short-range order in Ni-Au solid solutions.

(v) It is interesting that in the phase separating Ni-Au and Cu-Ag
the lowest interfacial energy $I(n \rightarrow \infty,\widehat{G})$ 
occurs for the close-packed \{111\} interfaces, and the highest for 
\{001\} interfaces. This situation is completely 
different in the ordering systems Cu-Au and Ag-Au, which have \{110\}
as the lowest and either \{111\} or \{001\} as the highest 
$I(n \rightarrow \infty,\widehat{G})$.

(vi) Figure~\ref{fig:SL} shows the enrgies of the random alloys at the
equiatomic composition. We see that in Cu-Au and Ag-Au all {\it
long-period\/} superlattices are unstable with 
respect to the formation of a random alloy. In Ni-Au the random
alloy is less favorable than coherent phase
separation in the $\langle 001 \rangle$ direction, 
but slightly more favorable than
infinite coherent superlattices along $\langle 110 \rangle$ and 
$\langle 111 \rangle$. However,
{\it short-period\/} $\langle 110 \rangle$ superlattices are lower in
energy than the random alloy. All $\langle 111 \rangle$ superlattices
of NiAu have higher formation enthalpies than the random alloy. In
Cu-Ag only the long-period 
$\langle 001 \rangle$
superlattices have lower bulk formation enthalpies than the random
alloy. The epitaxial growth of CuAg and NiAu alloys is discussed more
thoroughly in Sec.~\ref{sec:alloys}.

(vii) In size-mismatched systems (Cu-Ag, Cu-Au, and Ni-Au) 
$\Delta H_{\rm SL}(n,\widehat{G})$ exhibit the same order as the
constituent strain $\Delta E_{\rm CS}^{\rm eq}(x,\widehat{G})$, i.e.,
$\Delta H_{\rm SL}(n,[001])$ is lowest and $\Delta H_{\rm
SL}(n,[111])$ is highest. It suggests that low constituent strain
stabilizes even short-period superlattices.

\subsection{Comparison of the trends in stability of metal and
semiconductor superlattices}

Growth of semiconductor superlattices is a more mature are than
than growth of metal superlattices, and much more data are available
at present. Thus, it is of interest our results in
Figs.~\ref{fig:SL} and \ref{fig:interf} for metals with analogous
results for semiconductors.\cite{dandrea90,wei+AZ}

\begin{figure*}
\epsfxsize=6in
\centerline{\epsffile{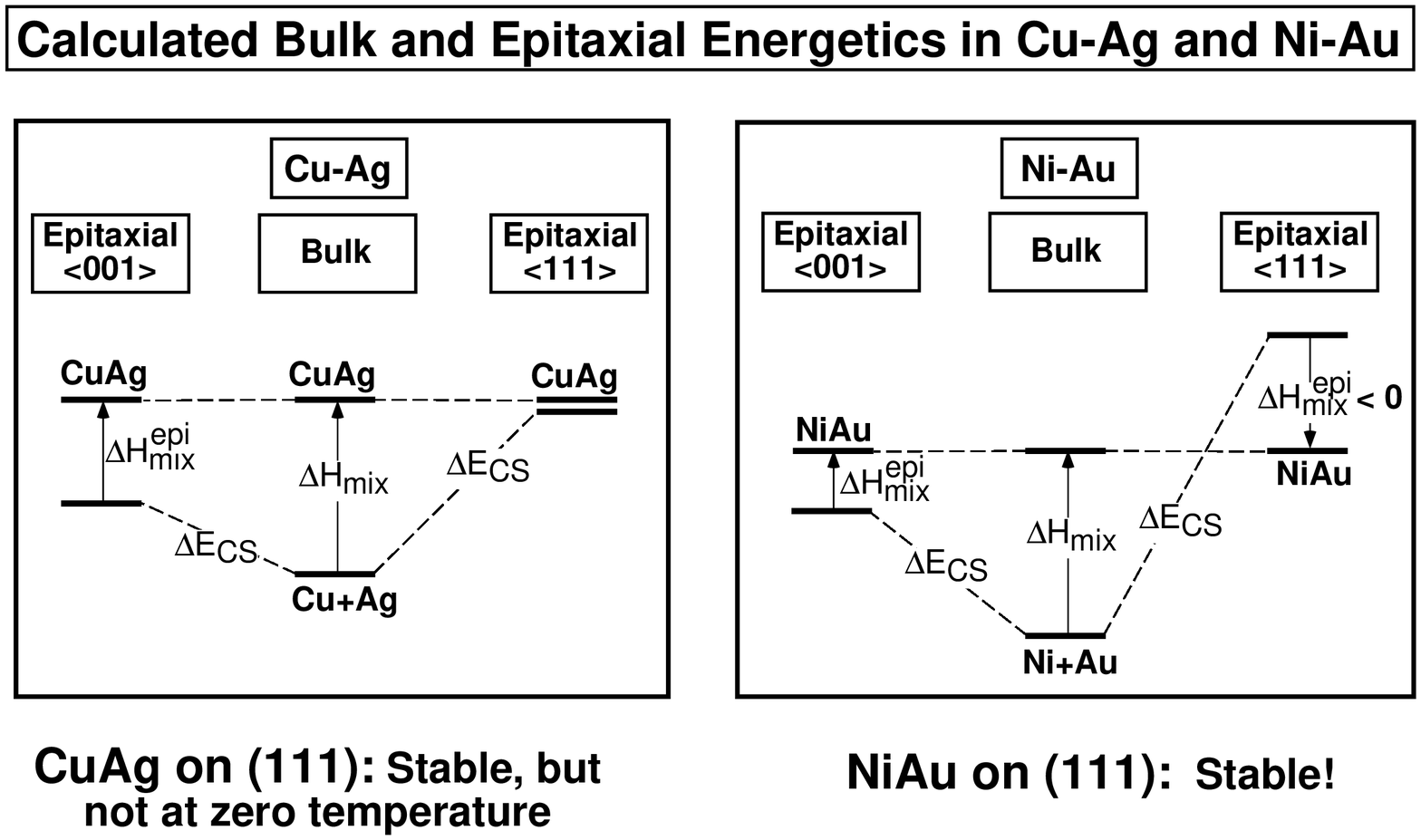}}
\vskip 5mm
\caption{Mixing enthalpies $\Delta H_{\rm mix}$ (in meV/atom) for bulk
and epitaxial equiatomic Cu-Ag and Ni-Au alloys. All epitaxial
calculations assume that the substrate is lattice matched to the
random alloy. $\Delta E_{\rm CS}$ is the sum of epitaxial strain
energies of pure elements [see Eq.~(\protect\ref{eq:epitaxialHmix})].}
\label{fig:Hmix}
\end{figure*}

Lattice-mismatched semiconductors generally have 
$\Delta H_{\rm mix}^{\rm bulk} (x) \geq 0$ and
$\Delta H_{\rm SL}^{\rm bulk} \geq 0$. Thus, they resemble Ni-Au and
Cu-Ag rather than the compound-forming system Cu-Au.
LDA calculations reveal that 
$\Delta H_{\rm SL}^{\rm bulk}(n,\widehat{G})$ for 
$G=\langle 111 \rangle$ and $G=\langle 001 \rangle$ look exactly like
in Cu-Ag or Ni-Au: the energy {\it decreases\/} as the period 
$n$ increases, and the interfacial energies are mostly
positive. However, in the $\langle 110 \rangle$ and 
$\langle 201 \rangle$ directions, the interfacial energies are 
{\it negative,\/} and $\Delta H_{\rm SL}^{\rm bulk}(n,\widehat{G})$
{\it increases\/} with $n$, like in Ni-Au and Cu-Au.
Hence, semiconductor superlattices behave generically as Ni-Au
superlattices. However, short-period $\langle 201 \rangle$ 
semiconductor superlattices (e.g., the chalcopyrite structure,
corresponding to $n=2$) have a {\it lower\/} energy than the random
alloy, while in Ni-Au it is the $\langle 001 \rangle$ short-period
superlattices that have lower energies than the random alloy.
Hence, while the Ni-Au random alloy can lower its energy by developing
$\langle 001 \rangle$ ordering, semiconductor random alloys can lower
their energy by developing $\langle 201 \rangle$ ordering.
Both in Ni-Au and semiconductor alloys, the ultimate ground state
is incoherent phase separation.

\subsection{Epitaxial growth and surface intermixing}
\label{sec:alloys}

Recent experimental studies\cite{danes,stevens95} have
grown epitaxial films of noble metal alloys which are immiscible
in the bulk form. For instance, Stevens and Hwang\cite{stevens95}
have grown Cu-Ag alloys on a Ru(0001) substrate, demonstrating that Cu
and Ag are miscible at $T=823$~K, despite the fact that in bulk, Cu
and Ag are strongly immiscible at this temperature and composition.
It has also been observed that Au deposited on Ni(110) surface
replaces it in the first surface layer forming a surface Ni-Au
alloy,\cite{danes} although Au is completely insoluble in bulk Ni.
In what follows we show that the stabilization of epitaxial
solid solutions from bulk-immiscible constituents 
can be explained by the additional {\it
destabilization\/} of the constituents due to the epitaxial
constraint. Indeed, Eq.~(\ref{eq:epitaxialHmix}) shows that the
epitaxial mixing enthalpy $\delta H_{\rm mix}^{\rm epi}$
may be considerably lower than the bulk mixing enthalpy
$\Delta H_{\rm mix}^{\rm bulk}$ if the sum of the constituent strain
energies on the right hand side is large.

Figure~\ref{fig:Hmix} shows the results for the epitaxial 
stabilization of equiatomic NiAu and CuAg alloys, assuming that
the substrate is lattice matched to the disordered alloy.

(i) Disordered CuAg and NiAu alloys have large positive
bulk mixing enthalpies $\Delta H_{\rm mix}^{\rm bulk}$,
in agreement with the observed bulk immiscibility.

(ii) Epitaxy destabilizes the constituents, and hence stabilizes the
epitaxial alloy in all cases. This effect is much larger for the 
elastically hard direction $\langle 111 \rangle$ than for the soft 
$\langle 001 \rangle$ direction.

(iii) The epitaxial mixing enthalpy $\delta H_{\rm mix}^{\rm epi}$
for $\langle 111 \rangle$ becomes {\it negative\/} in Ni-Au, showing
that the solid solution is energetically favored over the epitaxially
phase separated state. In CuAg,  $\delta H_{\rm mix}^{\rm epi}$ is still
positive and these alloys are unstable under epitaxial conditions
at $T=0$~K.

(iv) Epitaxial conditions lead to a significantly enhanced miscibility
since $\delta H_{\rm mix}^{\rm epi} \leq \delta H_{\rm mix}^{\rm bulk}$.
A simple mean-field estimate of the miscibility gap
temperature for CuAg grown on a nearly lattice-matched Ru(0001)
substrate [equivalent to a fcc(111) substrate]
gives $T_{\rm MG} = 2\Delta H_{\rm mix}^{\rm epi} = 150$~K. Thus,
for (111)-epitaxy at the temperature (823 K) of Steven's and Hwang's
experiment, our calculations predict complete solubility of Cu-Ag, as
observed. 

(v) The epitaxial stabilization is strongly dependent on the
substrate orientation. A bigger effect can be observed for elastically
hard directions, e.g., $\langle 111 \rangle$ and 
$\langle 110 \rangle$ for noble metal alloys.

\section{Summary}
\label{sec:summary}

We have investigated the effects of anharmonic strain on the stability
of epitaxial films, superlattices and epitaxially grown disordered
alloys. We find that anharmonic epitaxial strain produces certain
qualitative and quantitative features absent in the harmonic theory.
In particular,

(i) Epitaxial softening functions $q(a_s,\widehat{G})$ are strongly
dependent on the substrate lattice constant $a_s$, while they are
constants in the harmonic theory. For instance,
as a consequence of the small fcc/bcc and fcc/bct energy difference,
biaxially expanded Cu and Ni show drastic softening of $q(a_s,[001])$. 
Furthermore, biaxially compressed Cu, Ag, and Au have low values of
$q(a_s,\widehat{G})$ along directions $\langle 201 \rangle$ and 
$\langle 110 \rangle$ with relatively
loose packing of atoms in the epitaxial planes.

(ii) The dependence of $q(a_s,\widehat{G})$ on the direction
$\widehat{G}$ can differ from harmonic predictions. 
For instance, $\langle 110 \rangle$ is the hardest direction in
biaxially expanded Cu and Ni, and $\langle 201 \rangle$ is the 
softest in biaxially compressed Cu, 
Ag and Au. The harmonic formula always predicts either 
$\langle 111 \rangle$ as the hardest and $\langle 001 \rangle$ as the
softest direction, or vice versa.

(iii) The strain energy of infinite coherent superlattices exhibits
marked anomalies associated with the anharmonic behavior of
constituent $q(a_s,\widehat{G})$. The size-mismatched systems Cu-Ag,
Cu-Au and Ni-Au exhibit very low constituent strain for Ag- and
Au-rich  $\langle 001 \rangle$ superlattices, since $\langle 001
\rangle$  is the easy direction for biaxial
expansion of Cu and Ni. Similarly, $\langle 201 \rangle$ 
superlattices with small
Ag or Au content have low coherency strain energies because this is
the easy deformation direction for biaxially compressed Ag and Au.
The in-plane lattice parameter $a_{\rm SL}$ of long-period 
$\langle 001 \rangle$
superlattices suffers a discontinuous jump around $x \approx 0.2$, and
other directions show considerable deviations from linear behaviour.

(iv) These elastic anomalies are less pronounced in short-period
superlattices, although they contribute to the structural stability
of $\langle 001 \rangle$ superlattices. 
Short-period bulk superlattices are stable
in Ag-Au and Cu-Au due to negative interfacial energies. Ag-Au and
Ni-Au have positive interfacial energies, leading to superlattice
formation being energetically unfavorable with respect to phase 
separation. The interaction energy between interfaces in Ni-Au is so
strong that short-period ($n \propto 2$) superlattices along 
$\langle 110 \rangle$
are more stable than the long-period superlattices with fewer
interfaces.

(v) Epitaxially grown disordered alloys can be stabilized even if the
system phase separates in bulk form. This effect is caused by
additional destabilization of the phase separated state due to the
epitaxial constraint on the constituents, requiring them to be
coherent with the substrate. The stabilization is more pronounced for
elastically hard directions with high values of $q(a_s,\widehat{G})$,
e.g. $\langle 111 \rangle$. For instance, we find that even though 
Ni-Au and Cu-Ag
phase separate in the bulk ($\Delta H_{\rm mix}^{\rm bulk} (x) > 0$), 
equiatomic Ni$_{0.5}$Au$_{0.5}$ alloys are miscible when
grown on a lattice-matched (111) substrate, while Cu$_{0.5}$Ag$_{0.5}$
on a (111) substrate is immiscible at $T=0$~K but miscible at
$T>150$~K. Neither Ni$_{0.5}$Au$_{0.5}$ nor Cu$_{0.5}$Ag$_{0.5}$ 
are miscible when grown on a lattice-matched (001) substrate,
corresponding to a low energy penalty on the phase separated
constituents.

\acknowledgements

This work has been supported by the Office of Energy Research,
Basic Energy Sciences, Materials Science Division, U.S. Department of
Energy, under contract DE-AC36-83CH10093.


\begin{references}
\bibitem{explainBain}
Indeed, since $E'(\gamma) < 0$ for $\gamma \ll \gamma_{\rm bcc}$
and $E'(\gamma) > 0$ for $\gamma \gg \gamma_{\rm fcc}$, 
the derivative $E'(\gamma)$ must pass through zero an odd
number of times, proving that in addition to the two points of cubic
symmetry there must exist a third extremal point corresponding
to the bct structure. The implicit assumption that there are no
inflection points does not change the essence of the discussion.
\end{references}
\end{document}